\documentclass[a4paper, 12pt]{article}
\usepackage[left=3cm,right=2cm,top=1cm,bottom=2cm]{geometry}
\usepackage[utf8]{inputenc}
\usepackage[english]{babel}
\usepackage{graphicx}
\usepackage{graphics}
\usepackage[hidelinks]{hyperref}
\usepackage{authblk}
\author[1]{Ioanna A. Gorbunova}
\author[1]{Nikolai O. Bezverkhnii}
\author[1]{Maxim E. Sasin}
\author[1]{Idris M. Gadzhiev}
\author[1,*]{Oleg S. Vasyutinskii}
\affil[1]{Ioffe Institute, Polytekhnicheskaya 26, 194021 St.Petersburg, Russia}
{
       \makeatletter
       \renewcommand\AB@affilsepx{: \protect\Affilfont}
       \makeatother

       \affil[ ]{Email ids}

       \makeatletter
       \renewcommand\AB@affilsepx{, \protect\Affilfont}
       \makeatother

       \affil[*]{osv@pms.ioffe.ru}

}

\title{\textbf{Ultrafast polarization-modulation transient spectroscopy to study electronic excited state dynamics in solutions and cells}}
\date{}

\begin{document}
       \maketitle
\begin{abstract}
 A novel polarization-modulation transient method has been developed for studying fast anisotropic relaxation in electronic excited states of polyatomic and biologically relevant molecules under excitation with femtosecond laser pulses. The method is based on the modulation of pump beam polarization with a photo-elastic modulator and detection of an anisotropic contribution to the transient signal by a highly-sensitive demodulation balanced scheme. The method was tested on  aqueous solution of coenzyme NADH (nicotinamide-adenine-dinucleotide) pumped at 360~nm and probed at 720 nm. Anisotropic vibrational relaxation and rotational diffusion have been observed in the sub-picosecond time domain. The method significantly enhances the accuracy of transient measurements and allows for recording of high-quality signals at low energy (a nJ) pump pulses.
\end{abstract}

Transient absorption spectroscopy is widely used nowadays as a powerful tool for investigation of ultrafast photoinduced processes of electron and proton transfer, isomerization, and excited state dynamics. These studies have received considerable interest within the broad scientific community because the dynamics of elementary physical and chemical processes in the gas and condensed phases can be followed in real time by femtosecond pump-probe spectroscopy \cite{Berera09,Henriksen14,Zhu15,Fisher16}. Ultrafast transient absorption spectroscopy of biologically relevant molecules under excitation with femtosecond laser pulses allows to reach a previously inaccessible level of understanding of dynamics of intra- and intermolecular interactions including relaxation in excited states, nonradiative recombination, energy transfer, ets \cite{Cohen03,Ewing14,Corrales16,Heiner13,Heiner17}.



Excited state dynamics in biologically relevant molecules is closely related to fast anisotropic relaxation and rotation diffusion processes, that can be successfully studied by multiphoton polarized fluorescence spectroscopy \cite{Couprie94,Lakowicz97a,Vishwasrao06,Denicke10,Blacker13,Herbrich15,Sasin18}, or by polarization-resolved pump-probe ultrafast spectroscopy \cite{Fayer05,Fenn09,Tros15,Corrales16,Rumble19,Hunger19}. In both cases a linearly polarized pump pulse resonantly promotes ground state molecules to an excited electronic/vibrational state thus producing an anisotropic distribution of excited molecular axes/bonds that becomes isotropic as a function of time due to random orientational motion of the molecules and relaxation processes. Within the pumpe-probe procedure a real time anisotropy decay is measured by absorption change with a delayed probe pulse that is polarized either parallel, or perpendicularly to the pump pulse:
\begin{equation}
\label{eq:Idiff1}
R(\tau)=\frac{I_{\|}(\tau)-I_{\bot}(\tau)}{I_{\|}(\tau)+2I_{\bot}(\tau)},
\end{equation}
where $\tau$ is a delay time between the pump and probe pulses.

Anisotropic relaxation and rotation diffusion decay times that can be determined in these experiments are of great importance for understanding of complicated biochemical processes in living cells and tissues. For instance, the rotation diffusion time $\tau_{rot}$ is known to be proportional to the solution viscosity and therefore it can be used for determination of local intracellular and extracellular viscosity. Moreover, the decay time values depend on the location of the molecule inside or outside a cell and can be used as a noninvasive marker of intracellular processes and indirectly also of cellular energy metabolism \cite{Schaefer19}.

Major problems in measuring an absorption change $\Delta I_{ab}(\tau)=I_{\|}(\tau)-I_{\bot}(\tau)$ in cells is short decay time values (on the order of a ps) and a small signal-to-noise ratio because transient light intensities $I_{\|}(\tau)$ and $I_{\bot}(\tau)$ in eq.(\ref{eq:Idiff1}) transmitted through the experimental sample are  typically much larger than the absorption change $\Delta I_{ab}(\tau)$ that is proportional to the absorbing molecule concentration and to the pump pulse energy. This is especially true at relatively low pump pulse energy needed to prevent a cell damage by the laser light. Therefore, the experimental signal noise mainly caused by fluctuations in the intensity and beam pointing of the probe pulses.

The problem of decreasing the noise is currently solved for IR pump-probe experimental schemes by several methods. Fenn et al \cite{Fenn09} successively measured $I_{\|}(\tau)$ and $I_{\bot}(\tau)$ using a polarizer placed in the probe beam, and averaged many successive parallel and perpendicular measurements. Tros et al \cite{Tros15} suggested an improvement of the approach \cite{Fenn09} by measuring  $I_{\|}(\tau)$ and $I_{\bot}(\tau)$  alternately on a shot-to-shot basis. In this way, the IR fluctuations at much smaller frequencies than the pulse repetition rate (about 1~kHz \cite{Tros15}) effectively canceled out, resulting in a better signal-to-noise ratio. However, in that case a higher frequency noise still contributes to the measured anisotropy in eq.(\ref{eq:Idiff1}).

In this paper we present a novel polarization-modulation transient  (PMT) method that effectively damps the noise in the absorption change $\Delta I_{ab}(\tau)$ in the entire power spectrum and allows to achieve high-quality transient absorption signals at very low (a few nJ) pump pulse energies and with a sub-picosecond time resolution. A femtosecond oscillator with a pulse duration of about 100 fs, pulse repetition rate of 80 MHz and pulse energy of about 1 nJ was used in the experiment as a light source without any additional optical amplifier. The method is based on the modulation of the pump pulse train polarization at 100 kHz by means of a photo-elastic modulator with following  separation of the anisotropic contribution to the signal using a highly-sensitive balanced detection scheme, a differential integrator, and a lock-in amplifier recording the absorption change $\Delta I_{ab}(\tau)$ at the modulation frequency of 100 kHz in a very narrow frequency bandwidth of a few Hz.

The method significantly enhances the accuracy of  transient polarization-sensitive measurements. To the best of our knowledge the experimental result presented in this paper is the first to demonstrate the transient pump-and-probe spectroscopy in biological molecules at a nJ level of pump pulses energy.

The method was tested on the study of fast anisotropic relaxation and rotational diffusion in electronic excited states in biological coenzyme NADH (nicotinamide-adenine-dinucleotide) in aqueous solution. Coenzyme NADH is one of the most important biological molecules, since it takes an active part in redox reactions in living cells. Possessing intense fluorescence emission NADH is widely used as a natural fluorescence marker for studying biochemical processes in living cells.  Up to now, a number of studies of NADH fluorescence in solutions and cells have been carried out mostly dealt with time-resolved fluorescence measurements (see e.g. \cite{Freed67,Vishwasrao06,Sasin19,Vasyutinskii17,Blacker19}).

However, despite of numerous studies, energy transfer processes in the excited states of NADH are currently far from clear understanding. NADH has complicated excited state photodynamics possessing two chromophore groups: nicotinamide and adenine that can interact with each other under certain conditions \cite{Schaefer19}. This dynamics was investigated recently by Heiner et al \cite{Heiner13,Heiner17} using the transient absorption spectroscopy. No polarization-dependent measurements have been carried out till now.

\begin{figure}[htbp]
\centering
\includegraphics[scale=0.5]{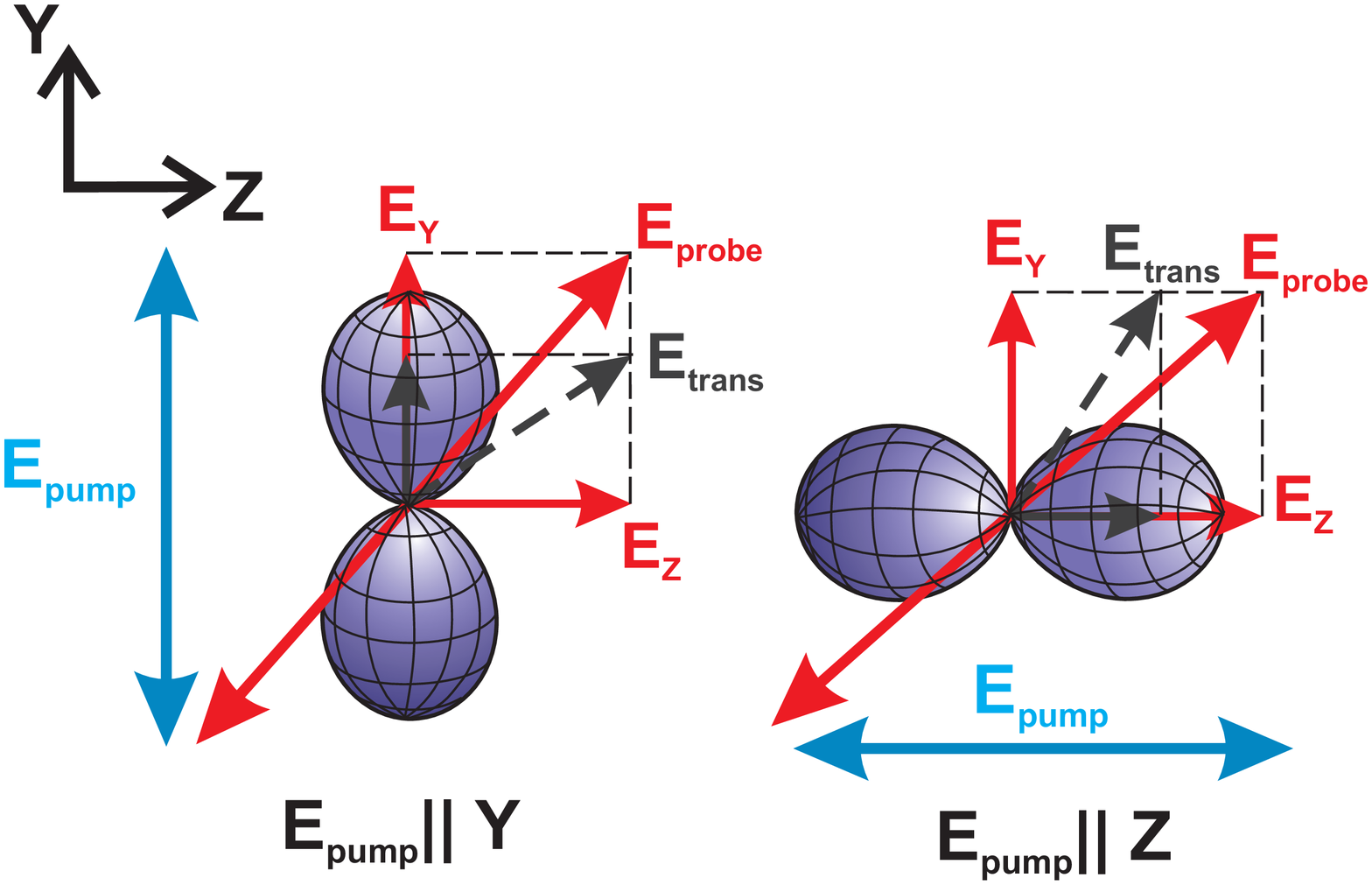}
\caption{Experimental method. $E_{pump}$ is a pump beam polarization, $E_{probe}$ is a probe beam polarization, $E_{trans}$ is the polarization of transmitted probe beam. Left side: the alignment of excited molecular axes at pump beam polarization along Y axis. Y probe beam component undergoes larger absorption by molecules than Z component. Right side: the alignment of excited molecular axes at pump beam polarization along Z axis. Z probe beam component undergoes larger absorption than Y component.}
\label{fig1}
\end{figure}

The experimental geometry used in this paper is shown schematically in Fig. \ref{fig1} where both pump and probe beams are linearly polarized and propagate collinearly along axis X. The pump beam at 360 nm was a frequency-doubled fundamental output from the femtosecond oscillator. The pump beam wavelength lay within the first NADH absorption band referred to the nicotinamide chromophore group and excited the NADH molecules from the ground electronic state to the first excited electronic state. The direction of the pump beam polarization oscillated between vertical (Y) and horizontal (Z) positions at the frequency of 100 kHz thus producing the anisotropy (alignment) distribution of excited molecular axes along Y and Z, respectively.

The  probe beam at 720 nm was the fundamental output of the same laser linearly polarized on 45 degrees to axis Z. This beam could be absorbed only from the molecular excited state and transferred the molecule to the higher laying electronic excited states related to the adenine chromophore group. No probe beam absorption occurred in the absence of the pump beam.  As can be seen in Fig. \ref{fig1} the absorption of the I$_Y=E^2_Y$ and I$_Z=E^2_Z$ probe beam components differed from each other depending on the pump beam polarization.

Therefore, the I$_Y$ and I$_Z$ polarization components of the probe beam transmitted through an absorption cuvette oscillated in quadrature to each other with a frequency of 100 kHz. These components were separated by a polarization prism and then recorded simultaneously and independently by two photodiodes.  Electric signals from the photodiodes entered two inputs of a differential integrator with a pass band of 0--4 MHz. Finally the  differential integrator output modulated at 100 kHz was directed to a high-frequency lock-in amplifier where the differential signal $I_{diff}(t,\tau)$ at the 100 kHz modulation frequency was detected and averaged.

The signal $I_{diff}(t,\tau)$ after the differential integrator and before the lock-in amplifier can be presented in the form:
\begin{equation}
\label{eq:Idiff2}
I_{diff}(t,\tau) = -\frac{I_0L}{2}(k_{\parallel}(\tau)-k_{\perp}(\tau))cos\delta(t),
\end{equation}
where $I_0$ is an initial probe beam intensity, L is a length of the absorption cuvette,  $k_{\parallel}(\tau)$ and $k_{\perp}(\tau)$ are absorption coefficients for the probe beam components polarized parallel and perpendicular to the pump polarization direction, respectively, and $\delta(t)$ = $\pi \sin(\Omega t)$ is a phase shift introduced to the pump beam by the photo-elastic modulator operated at $\Omega$=50 kHz.

The output of the lock-in amplifier was proportional to the absorption change signal: $\langle I_{diff}(t,\tau)\rangle \propto \Delta I_{ab}(\tau)$, where the angle brackets mean synchronous detection at the modulation frequency $2\Omega$. In fact, the recorded signal was due to the linear dichroism of the NADH solution induced by linearly polarized pump light. Determination of the anisotropic relaxation decay times in excited NADH was provided by tuning the delay time $\tau$ between the pump and probe pulses.

The main advantage of the suggested method is that the noise in the entire power spectrum can be effectively damped by the balanced detection scheme and by synchronous detection of the differential signal $I_{diff}(t,\tau)$ amplitude-modulated at 100 kHz in a very narrow frequency bandwidth of a few Hz.

\begin{figure}[htbp]
\centering
\includegraphics[scale=0.5]{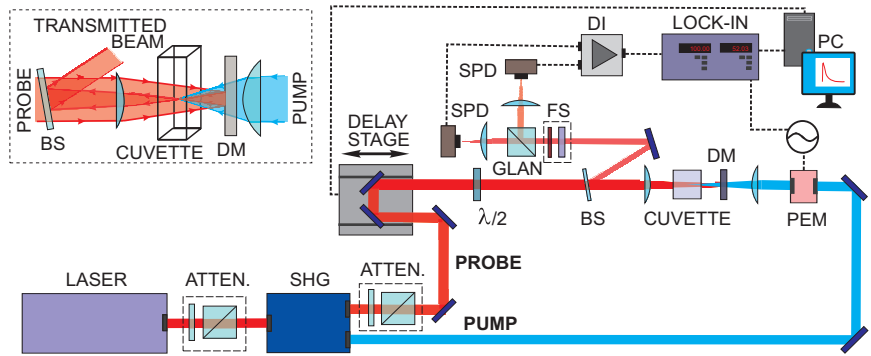}
\caption{Experimental setup. SHG is a second harmonic generator, PEM is a photoelastic modulator, BS is a beam splitter, DM is a dichroic mirror, FS is a filtering system, SPD are silicon photo diodes, DA is a differential integrator.}
\label{fig2}
\end{figure}

The experimental setup is shown in Fig \ref{fig2}. In breif, the fundamental output of the Ti:Sa oscillator (MaiTai, Spectra Physics) at 720 nm was frequency-doubled by a second harmonic generator (SHG) (Inspire Blue, Spectra Physics) and then used as a pump beam at 360 nm. After SHG the pump beam passed through a photo-elastic modulator (PEM) (PEM-100, Hinds Instruments) that operated at 50 KHz and modulated the plane of the pump beam  polarization from Y to Z and back at the double frequency of 100 kHz. Then the pump beam was focused through a dichroic mirror (DM) to the center of a quartz absorption cuvette of 10 mm length containing aqueous solution of NADH. Average pump beam intensity focused on the cuvette was about 80 mW. $\beta$-NADH (reduced disodium salt hydrate, Sigma-Aldrich) dissolved in the Tris buffer (hydroxy-methyl aminomethane hydrochloride) at the concentration of 0.1 mM was used in experiments. NADH solutions were prepared fresh daily. To prevent NADH bleaching the solution was slow circulated through the cuvette by a peristaltic pump.

The fundamental output of the laser at 720 nm was used as a probe beam. The probe beam passed successively through an attenuator and a motorized delay stage producing the probe pulse delay for the time $\tau$ with respect to the pump pulse. The probe pulse polarization plane was fixed at 45 degrees to Z axis by a half-wave plate as shown in Fig. \ref{fig1}.  As shown in inset in Fig. \ref{fig2} the probe beam passed through the absorption cuvette and then was reflected back by DM and focused onto the center of cuvette. The focal regions of the beams overlapped considerably only when the probe beam was reflected back from the DM, therefore the beams interacted with NADH molecules mainly when they propagated inside the absorption cuvette in parallel to each other. Average power of the probe beam on the cuvette was about 40 mW.

The transient probe beam exited the cuvette was reflected by a beam-splitter (BS) and then directed to a Glan prism (GLAN) for separating two orthogonal polarization components $I_{\|}(\tau)$ and $I_{\bot}(\tau)$. To avoid the pump light scattering into the registration channel, a filtration system (FS) consisted of a dichroic mirror and an absorption filter was placed in front of the Glan prism. The polarization components  $I_{\|}(\tau)$ and $I_{\bot}(\tau)$ were separated by the Glan prism, and then recorded by a balanced detection scheme consisted of two identical silicon photodiodes (SPD) (DET10A/M, ThorLabs) and a differential integrator (DI) with a pass band of 0--4~MHz. The balanced detection scheme was adjusted to have zero output electric signal when the pump beam was tuned off and therefore effectively cancelled fluctuations in the probe beam amplitude and in the solution density. The output differential signal of the DI  amplitude-modulated at 100~kHz was recorded and averaged by the lock-in amplified (LOCK-IN) (SR844 RF, Stanford Instruments) with a narrow frequency bandwidth of 3~Hz.

\begin{figure}[htbp]
\centering
\includegraphics[scale=1]{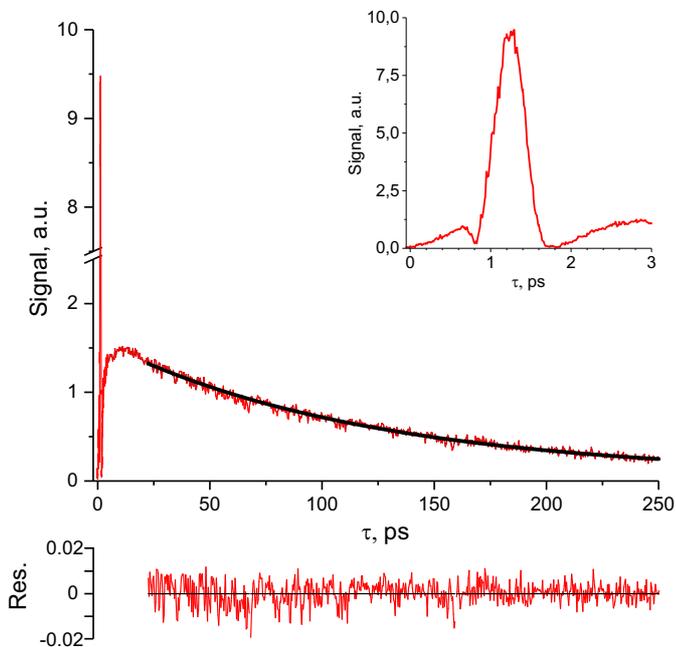}
\caption{Experimental signal: red curve represents the experimental data and black curve is a fit. Inset shows a signal fragment at short delay times in an enhanced scale}
\label{fig3}
\end{figure}

A typical experimental signal is presented in Fig.~\ref{fig3} with a red curve. As can be seen in Fig.~\ref{fig3} at small delay times the signal has a high narrow pick that is shown on an extended scale in the inset. We suggest that the pick reflects complicated multiphoton absorption processes occurring in NADH and water during the time of overlap between the pump and probe pulses. The pick has full width half maximum (FWHM) of 0.6~ps that is likely a result of contribution of each pulse duration of about 0.1~ps and the pulses dispersion inside the absorption cuvette.

At larger delay times (approximately $\tau\geq$2~ps) the signal in Fig.~\ref{fig3} reflects the dynamics of anisotropic vibrational and rotational relaxation in NADH molecules. This part of the signal did not exist in pure water.  To the best of our knowledge the anisotropic vibrational relaxation within a few picoseconds delay time under excitation via the first NADH absorption band has never been observed earlier. The interpretation of the shape of this part of the signal is still under study in our laboratory. The decaying part of the anisotropy signal at $\tau \geq$25~ps in  Fig.~\ref{fig3} can be explained by the rotation diffusion of excited NADH molecules in solution.

The results of recent time-resolved polarized fluorescence studies on NADH in solutions \cite{Vishwasrao06,Blacker13,Vasyutinskii17,Sasin19,Blacker19} suggest that the fluorescence intensity decays multiexponentially with two isotropic decay times $\tau_1$ and $\tau_2$ and one, or two rotation diffusion times. The nature of the two lifetimes in NADH is still not clear now. Couprie et al \cite{Couprie94} argued that different lifetimes characterise folded and unfolded conformers of NADH, while Blacker et al \cite{Blacker19} suggested that both times are internal characteristic of the nicotinamide molecular group. As reported by Vasyutinskii et al \cite{Vasyutinskii17} under excitation of NADH molecules in aqueous solution by femtosecond laser pulses via the first absorption band the fluorescence intensity decays tri-exponentially with the times $\tau_1$=220$\pm$ 20~ps and $\tau_2$=710$\pm$ 20~ps, and a single rotation diffusion time  $\tau_{rot}$=170$\pm$ 20~ps.

A simplified theoretical expression describing the transient signal decay in Fig.~\ref{fig3} at delay times $\tau$ > 25~ps can be presented in the following form:

\begin{equation}
\label{eq:theor}
\Delta I_{ab}(\tau) \propto \left[a_1e^{-\tau/\tau_1}+a_2e^{-\tau/\tau_2}\right]r_0\,
e^{-\tau/\tau_{rot}},
\end{equation}
where $r_0$ is the anisotropy \cite{Lakowicz97a} and $a_1$, $a_2$ are weighting coefficients ($a_1$ + $a_2$ =1).

We fitted the experimental data in Fig.~\ref{fig3} using eq.(\ref{eq:theor}) where the values $\tau_1$, $\tau_2$, $a_1$, and $a_2$ were fixed and taken from ref.~\cite{Vasyutinskii17}, while the values $\tau_{rot}$ and $r_0$ were used as fitting parameters. A corresponding fitting curve is shown in Fig.~\ref{fig3} in black. The rotation diffusion time determined from the fit was found to be $\tau_{rot}$= 218 $\pm$ 15~ps that is in a reasonable agreement with the value of 170$\pm$ 20~ps obtained earlier by Vasyutinskii et al \cite{Vasyutinskii17} using time-resolved polarized fluorescence method.

Note that the temporal resolution of the fluorescence method \cite{Vasyutinskii17} was strongly limited by the response time of the photodetectors used that was about 200~ps, while the temporal resolution of this paper PMT method is about 0.6~ps as can be seen in Fig.~\ref{fig3}. Therefore, the PMT method demonstrates much better temporal resolution than the polarized fluorescence method and can provide detailed information on the anisotropic vibronic relaxation and rotation diffusion in excited states of biologic molecules in the sub-picosecond time domain.

Therefore, in this paper a novel polarization-modulation transient method has been developed for investigation of anisotropic vibronic relaxation and rotation diffusion in excited states of biologic molecules. The method is based on the modulation of the pump pulse train polarization at 100 kHz with a photo-elastic modulator and with following  separation of the anisotropic contribution to the signal using a highly-sensitive balanced detection scheme and a lock-in amplifier recording the absorption change at the modulation frequency in a narrow frequency bandwidth of a few Hz. The method significantly enhances the accuracy of  transient polarization-sensitive measurements as it effectively damps the noise due to the fluctuations in the probe beam and in absorption solution in the entire power spectrum.

The method was tested on the study of fast anisotropic relaxation and rotational diffusion in electronic excited states in biological coenzyme NADH in aqueous solution under excitation at 360~nm and demonstrated the temporal resolution of 0.6~ps with a very low pump pulse energy of a few nJ. The experimental signals contain contributions from anisotropic vibrational relaxation within a few picosecond time delay and rotation diffusion within larger delay times $\tau \geq$25~ps. The anisotropic vibration relaxational has never been observed before in this excited state. The determined rotation diffusion time is in a good agreement with that obtained before by polarised fluorescence method.
\section*{Acknowledgments}
The study was supported by the Russian Foundation for Basic Researches under the grant No 18-03-00038a and by the BASIS foundation, grant No 19-1-1-13-1.

\bibliography{pump-probe_2020-1}
\end{document}